\documentclass[journal]{IEEEtran}

\usepackage{graphicx}
\usepackage{epstopdf}
\usepackage{amsmath}
\usepackage{amssymb}
\usepackage{amsthm }
\usepackage{epsfig}
\usepackage{times}
\usepackage{setspace}
\usepackage{bm}
\usepackage{siunitx}
\usepackage{threeparttable}
\usepackage{balance}
\usepackage{tabularx}
\usepackage{subcaption}
\usepackage{xcolor}
\usepackage{fancyhdr}
\usepackage{cite,url}

\usepackage{array}

\usepackage[T1]{fontenc}
\usepackage{times}
\usepackage[mathscr]{eucal}
\usepackage{wrapfig}
\usepackage{multirow}
\usepackage{tablefootnote}
\usepackage{makecell}

\newcommand{\equals}{\!=\!}
\newcommand{\lteq}{\!\le\!}
\newcommand{\minus}{\!-\!}
\newcommand{\gteq}{\!\ge\!}
\newcommand{\gthan}{\!>\!}
\newcommand{\lthan}{\!<\!}
\newcommand{\plus}{\!+\!}

\newcommand{\F}{\mathrm{F}}

\newcommand{\e}{\mathrm{e}}

\newcommand{\E}{\mathrm{E}}
\newcommand{\dx}{\:\mathrm{d}}

\newtheorem{lemma}{Lemma}

\begin{document}

% \title{UAV-Aided Data Collection in LoRa Sensor Networks Enabled by Wake-Up Radios}

% \title{A Low-Complexity Transmission Scheme for UAV-Aided LoRa Sensor Networks Featuring Wake-Up Radios}

% \title{Time-Constrained Erasure Correction for Data Recovery in UAV-LoRa-WuR Networks}
\title{Time-Constrained Erasure Correction for Data Recovery in UAV-Aided IoT}

\author{

Kushwanth Sistu and Siddhartha S. Borkotoky \vspace{-8mm}

\thanks{
This work was supported by the Department of Science and Technology (DST), Government of India, through its Core Research Grant (CRG) No. CRG/2023/005803. The authors are with the School of Electrical and Computer Sciences, Indian Institute of Technology Bhubaneswar, Khordha 752050, India (e-mail: 20ee01042@iitbbs.ac.in, \mbox{borkotoky}@iitbbs.ac.in)}

}
\maketitle
\thispagestyle{fancy}
\lhead{{\color{gray} This work appears in the IEEE Communications Letters (DOI: 10.1109/LCOMM.2025.3633952). This is the author's version, and posted for personal use, not for redistribution.}}

\begin{abstract}
We propose data-recovery schemes to improve the uplink reliability of Internet of Things (IoT) networks in which a  gateway mounted on an unmanned aerial vehicle (UAV) collects data from sensor nodes on the ground.  The proposed techniques employ fountain coding and message replication to correct frame erasures on the sensor-to-UAV links. 
A salient feature of the proposed schemes is their adaptive selection of the redundancy amount depending on the contact times between a sensor and the UAV. We illustrate our design philosophy for the special case of a network in which wake-up radios (WuRs) are used for activating the sensors upon the UAV's arrival, and the sensor-to-UAV data transfer employs the Long Range (LoRa) communication technology with random access.    
\end{abstract}


\vspace{-4mm}
\section{Introduction}
\label{sec:Intro}
UAVs as data aggregators in  IoT networks can provide tremendous advantages in terms of device energy consumption, coverage, and flexibility~\cite{ABM24,ZhC22,BBG18}. In a typical UAV-aided IoT scenario, a gateway mounted on a UAV hovers over a cluster of sensors from time to time. The sensors transmit their messages to the UAV, which relays them over a high-capacity link to the sink. The messages may represent sensor readings of physical parameters~\cite{BBG18}, machine-learning model parameters for federated-learning applications~\cite{NDP21}, etc. Since sensor-to-UAV links are typically lossy due to low transmission power and interference~\cite{XLX23}, data-recovery mechanisms may be required on these links to meet the desired reliability levels of the IoT application.

Frame-level erasure-correction schemes such as message replication and fountain coding have been successfully used for data recovery in IoT networks. These approaches transmit redundant messages to compensate for lost messages. Their incorporation into UAV-aided IoT, however, requires additional considerations. For example, the limited hovering time of the UAV  restricts the amount of redundant messages a sensor can transmit. Furthermore, since all sensors in a cluster compete for transmission times within the same UAV hovering window, excessive redundancy transmission may in fact be counterproductive due to increased collisions. These considerations must be incorporated into the design to effectively leverage the benefits of erasure correction in UAV-IoT. To our knowledge, existing literature lacks such studies, although prior work has demonstrated the reliability improvements offered by erasure coding when the contact time between a sensor and the data sink is short~\cite{DML16}. However,~\cite{DML16} does not consider the multipoint-to-point communication scenario that arises when a UAV collects data from a sensor cluster, and the constraints imposed on a  sensor's transmission policy by the potentially variable hovering time and the sensor's energy budget. Another approach toward improving the reliability of sensor-to-UAV message transfer is to eliminate frame collisions by using TDMA~\cite{SAA25}. However, TDMA requires a reliable connection setup phase, in which the UAV learns which sensors wish to transmit and the size of their respective data and accordingly assigns time slots to each.  Random access protocols avoid this requirement but increase collisions and losses. 
 
To avoid the complexity of TDMA, this letter focuses on a random-access scenario amenable even to unsophisticated IoT nodes and develops methods for improving communication reliability. Specifically, we contribute the following: 
\begin{itemize}
    \item We devise two erasure-correction schemes to improve the reliability of message transfer from a cluster of sensors to a UAV gateway.  One scheme employs fountain coding, and transmits redundant frames that are linear combinations of the sensor messages. The other scheme transmits message replicas. To our knowledge, this is the first investigation of this kind.

    \item Unlike conventional erasure correction, our schemes adaptively determine the amount of redundancy to be sent depending on the available contact time between the sender and the UAV. Consequently, they can adapt to scenarios where the UAV's hovering times are variable. 
    % In fact, our design philosophy can be extended to other mobile IoT scenarios beyond UAV-IoT. 
    
    \item  We illustrate our design for a practical UAV-IoT setup in which the sensors and the UAV are equipped with wake-up radios (WuR) and Long Range (LoRa) transceivers. 
    
    \item We develop analytical models for the schemes' performance to aid in system design.
\end{itemize}

\vspace{-3mm}
\section{System Model}
\label{sec:sys_mod}
\vspace{-1mm}
We consider a cluster of $n$ sensors or end devices (EDs). Each ED contains a LoRa transceiver as the main radio. In addition, it has a WuR. During idle times, the EDs keep their main radio switched off to conserve power. The WuR, which consumes negligible power~\cite{PMK17}, is kept on at all times. The UAV occasionally visits the cluster and hovers above it for a certain duration.  A sensor has $\beta$ messages to send to the UAV within the hovering time. For channel access by the EDs, we employ the protocol of~\cite{PaB23}, which uses
slotted communications from the EDs to the UAV with slot length $T_s$. The UAV hovers for $N_s$ slots, numbered 0 through $N_s \minus 1$. At every slot beginning, the UAV broadcasts a wake-up call (WuC) whose data field contains the current slot number $i$ and the value of $N_s$. Once awake, an ED transmits its messages to the UAV using one of the methods to be described in Section~\ref{sec:scheme}. Since each WuC serves as a time reference, we do not assume the use of additional synchronization mechanisms. However, techniques such as~\cite{HCW24} may be employed if desired.

For every transmission on the ED-to-UAV uplink, one among $N_f$ orthogonal frequency bands is chosen uniformly at random. Likewise, the spreading factor (SF) of the LoRa signal~\cite{GeR17} is chosen uniformly at random from the set $\{7,8,\ldots,k_m\}$, where $7 \lteq k_m \lteq 12$. (The SF specifies the number of bits carried per LoRa modulation symbol. Commercial LoRa modules typically support SFs 7 through 12). We assume that a sensor is capable of transmitting a maximum of $N_{\max}$ frames per data-collection cycle, where $N_{\max} \gthan \beta$. Typically, the energy profile of a sensor determines $N_{\max}$. An example calculation of $N_{\max}$ is given in Appendix~\ref{App:Nmax}.

\section{Redundancy Transmission Schemes}
\label{sec:scheme}
Our  schemes transmit up to \mbox{$\beta \plus \varepsilon$} frames, where the design parameter \mbox{$\varepsilon \lteq N_{\max} \minus \beta$}  represents the  amount of redundancy sent for correcting frame erasures. We define
$N(i) \equals N_s \minus i$
and 
$\gamma(i) \equals N(i)  \minus \beta.$
Thus, for an ED that wakes up in slot $i$, there are $N(i)$ slots remaining in the session, 
of which $\gamma(i)$ 
are available for redundancy transmission. Note that $\gamma(i)$ may be 0 or negative depending on the value of $i$. 
The two erasure-correction schemes operate as follows:

\vspace{-2mm}
\subsection{Fountain Coding}
An ED first checks if there are enough slots available to transmit $\varepsilon$ redundant frames, that is, whether $\gamma(i) \gteq \varepsilon$. 

If $\gamma(i) \gteq \varepsilon$: The ED encodes the $\beta$ source messages into \mbox{$\beta \plus \varepsilon$} coded messages and transmits them over $\beta \plus \varepsilon$ randomly chosen slots out of the $N(i)$ available slots. Each coded message is a weighted linear combination of the source messages. For the random linear fountain code~\cite{RLL22} used in our illustration, the $j$th coded message is given by
$ c_j \equals \sum_{l=0}^{\beta-1} \alpha_l^{(j)} \psi_l$, where $\psi_l$, $0 \leq l \leq \beta \minus 1$ denote the source messages and the coefficients $\alpha_l^{(j)}$ are chosen at random from a Galois field of a given order $q$ (GF($q$)). Upon receiving a set of $z$ coded messages that contain $\beta$ linearly independent combinations of the source messages, the receiver solves a system of linear equations (e.g., via Gaussian elimination) to obtain the source messages. By contrast, if fewer than $\beta$ linearly independent combinations are received, then fountain decoding fails. The probability that fountain decoding succeeds upon receiving $z$ coded messages is given by $P_{\mathrm{dec}}(z) \equals \prod_{v=0}^{\beta-1}(1-q^{v-z})$ for $z \gteq \beta$, and 0 for $z \lthan \beta$~\cite{RLL22}.  

If $\gamma(i) \lthan \varepsilon$: Fountain coding is not applied. Instead, if $\beta \lteq N(i)$, the $\beta$ messages are transmitted over $\beta$ randomly chosen slots. If $\beta \gthan N(i)$, then the ED randomly selects $N(i)$ messages and sends them over the $N(i)$ slots. Messages not selected are discarded and deemed lost.

\vspace{-2mm}
\subsection{Message Replication}

This scheme makes transmission decisions as follows:

$\gamma(i) \gteq 0$: Let
$
\hat{\varepsilon}(i) \equals \min\{\gamma(i),\varepsilon\},    
$
and $m_q$ and $m_r$ be the quotient and the remainder, respectively, when $\hat{\varepsilon}(i)$ is divided by $\beta$. 
Each message is sent $m_q \plus 1$ times; further, $m_r$ messages are randomly picked for one additional transmission. %Thus, a total of $\hat{\varepsilon}(i)$ frames are transmitted. 

$\gamma(i) \lthan 0$: This implies $N(i) \lthan \beta$. The ED randomly selects $N(i)$ messages and sends them over the $N(i)$ slots. Messages not selected are discarded and deemed lost.

\section{Performance Analysis}
\label{sec:analysis}
\vspace{-1mm}

Let the \textit{message delivery probability} denote the  probability that an arbitrary message from an ED is delivered to the UAV. 
Let $P_b$ be the probability that an ED successfully receives a WuC from the UAV. Then, the probability mass function for the slot in which an ED wakes up (denoted $W$) is 
\begin{align}
    P_W(i)  =(1-P_b)^{i} P_b         
\end{align}
for $0 \leq i \leq N_s-1$, and 0 otherwise.

\begin{lemma}
An ED that wakes up in slot $i$  transmits a frame to the UAV in slot $s$ with probability
\begin{align} \label{eq:cond_tx_prob_FC}
    P_{\mathrm{tx}}(s|W=i) =
    \begin{cases}
        (\beta + r)/N(i), \quad &s \geq i, \gamma(i) \gteq y \\
        \min\left\{\beta/N(i), 1\right\}, \quad &s \geq i, \gamma(i) \lthan y \\
        0, \quad &s < i,
    \end{cases}
\end{align}
where $r \equals y \equals \varepsilon$ for fountain coding and $r \equals \hat{\varepsilon}(i)$ and $y \equals 0$ for message replication.
\end{lemma}

\begin{IEEEproof}
    First, consider fountain coding. By definition, the ED has $N(i)$ slots available for use. If $\gamma(i) \gteq \varepsilon$, then $\beta \plus \varepsilon$ coded messages are randomly distributed over $N(i)$ slots; hence the probability of choosing a certain slot is $(\beta \plus \varepsilon)/N(i)$. If $\gamma(i) \lthan \varepsilon$, then $\beta$ uncoded messages are to be distributed over $N(i)$ slots; the probability of choosing a certain slot is either $\beta/N(i)$ (if $\beta \lteq N(i)$), or 1 (if $\beta \gthan N(i)$, meaning all slots are used and some messages are dropped.) This is succinctly expressed as $\min\{\beta/N(i),1\}$. 
    %Of course, the probability of choosing slot $s$ is 0 if $s \lthan i$.  
    For message replication, the proof is identical, except that an ED transmits redundancy for any $\gamma(i) \gthan 0$, and $\hat{\varepsilon}(i)$ is the number of redundant frames sent. 
\end{IEEEproof}

Let the \textit{transmission success probability} (denoted $\zeta(s)$)  be the probability that a frame sent in slot $s$ is successfully received at the UAV. Consider such a frame $\F$ from an arbitrary ED $\E$. Then $\zeta(s)$ is the probability that no other ED transmits in slot $s$ using the same frequency band with enough received power to cause the loss of $\F$. It can be determined as follows:

\begin{lemma}
The probability that a given  ED $\E'$ transmits a frame in a given slot $s$ is 
\begin{align} \nonumber 
\label{eq:coll_prob_FC}
P_{\mathrm{col}}(s) 
&=  \sum_{j=0}^{\min\{N_s-\beta - y,s\}}\frac{\beta + r}{N(j)} P_W(j) \\
&+ \theta_s \sum_{j= N_s-\beta - y +1}^{s}\min\left\{\frac{\beta}{N(j)}, 1\right\} P_W(j),
\end{align}
where $r \equals y \equals \varepsilon$ for fountain coding, $r \equals \hat{\varepsilon}(i)$ and $y \equals 0$ for message replication, and 
\begin{align}
    \theta_s =
    \begin{cases}
        1, \quad &s > N_s-\beta -y\\
        0, \quad &\text{otherwise}.
    \end{cases}
\end{align}

\end{lemma}

\begin{IEEEproof}
We first consider fountain coding. Suppose $\E'$ wakes up in slot $j$. Then $\gamma(j) \gteq \varepsilon$ if $j \lteq N_s \minus \beta \minus \varepsilon$. From the argument in Lemma 1, the probability of selecting slot $s$ in this case is $(\beta \plus \varepsilon)/N(j)$. The first term in~\eqref{eq:coll_prob_FC} averages this probability over all $j$ such that $\gamma(j) \gteq \varepsilon$. The $\min$ operation in the upper limit of the summation ensures that we consider only the values of $j$ up to $s$ (since an ED waking up after slot $s$ cannot transmit in slot $s$.) Similarly, the second summation accounts for the transmission probabilities for all $j$ such that $\gamma(j) \lthan \varepsilon$. The multiplier $\theta_s$ ensures that we consider the second summation only when there exists a $j$ that simultaneously satisfies both $\gamma(j) \lthan \varepsilon$ and $j \lteq s$.

For message replication, the proof is identical, except that an ED transmits redundancy for any $\gamma(i) \gthan 0$, and $\hat{\varepsilon}(i)$ is the number of redundant frames sent. \end{IEEEproof}

To derive the transmission success probability, consider frame $\F$ from ED $\E$ and an interfering frame $\F'$ from ED $\E'$. The probability that $\F$ and $\F'$ collide over the same channel is $P_{\mathrm{col}}(s)\mathcal{F}/N_f$.
Hence, the probability that one given ED (namely, $\E'$) enforces the loss of $\F$ is
\begin{align}  
    \mathcal{L}(s) = P_{\mathrm{col}}(s)\mathcal{F}/N_f.  
\end{align}
where $\mathcal{F}$ is the probability that $\F'$ is received with enough power to cause the loss of $\F$. An expression for $\mathcal{F}$ is derived in Appendix~\ref{App:FLP} for links with Nakagami-$m$ fading, a common model for ground-to-UAV links~\cite{YYF22}. Since there are $n \minus 1$ potential interferers, the transmission success probability for $\F$ (using the strongest interferer model~\cite{GeR17}) is
\begin{align} \label{eq:TSP}
    \zeta(s) = (1- \mathcal{L}(s))^{n-1}  
    = \left(1- P_{\mathrm{col}}(s)\mathcal{F}/N_f\right)^{n-1}. 
\end{align}

\vspace{-4mm}
\subsection{Message-Delivery Probability for Fountain Coding}
\label{sec:FC_analysis}

Consider an arbitrary message $\psi$ from an arbitrary ED $\E$ that woke up in  slot $i$. Then the following events can occur:
\subsubsection{$\gamma(i) \gteq \varepsilon$} In this case, $\E$ performs fountain coding and transmits $\beta \plus \varepsilon$ coded messages. Thus, the MDP for $\psi$ is the probability that the received set of coded messages is decodable. 
The message delivery probability then becomes
\begin{align}
    \mathcal{S}_1(i) \equals \sum_{z=\beta}^{\beta+\varepsilon} \xi(z,i)P_{\mathrm{dec}}(z)  \equals \sum_{z=\beta}^{\beta+\varepsilon} \xi(z,i) \prod_{v=0}^{\beta-1}(1 \minus q^{v-z}),
\end{align}
where $ \xi(z,i)$ is the probability that $z$ out of the $\beta \plus \varepsilon$ transmissions are successful. Finding an exact expression for $ \xi(z,i)$ is complicated due to the slot-dependent nature of the transmission success probability $\zeta(s)$. However, a fairly accurate approximation may be obtained by assuming that the transmission success events over different slots are mutually independent and have a fixed probability $\hat{\zeta}$, and then using the binomial mass function
\begin{align}
    \xi(z,i) \approx {\beta+\varepsilon \choose z} \hat{\zeta}^{z} (1 - \hat{\zeta})^{\beta+\varepsilon-z}.
\end{align}
The parameter $\hat{\zeta}$ is obtained by averaging the transmission success probabilities for the $N(i)$ slots that are available to the ED. That is 
\begin{align}
\label{eq:avg_zeta}
    \hat{\zeta} = \frac{1}{N(i)}\sum_{s=i}^{N_s-1} \zeta(s).
\end{align}

\subsubsection{$\gamma(i) \lthan \varepsilon$} 
In this case, fountain coding is not performed, and the message $\psi$ is sent as it is. 
The probability that $\psi$ is sent over slot $s$ is given by~\cite{PaB23}
\begin{align} \label{eq:tx_prob_2}  
    \mathcal{T}(s,i) =   \frac{1}{N(i)}  \min\left\{\frac{N(i)}{\beta},1\right\}, \quad i \leq s \leq N_s-1.
\end{align} 
The message delivery probability is thus
\begin{align}   \label{eq:suc_prob_2} 
\mathcal{S}_2(i) &= \sum_{s=i}^{N_s-1} \mathcal{T}(s,i) \zeta(s).
\end{align}

The overall message delivery probability is thus
\begin{align}
\label{eq:S_FC}
    \mathcal{S} = \sum_{i=0}^{N_s-\beta-\varepsilon}P_W(i)\mathcal{S}_1(i) + \sum_{i=N_s-\beta-\varepsilon+1}^{N_s-1}P_W(i)\mathcal{S}_2(i),
\end{align}
where the first summation is over those values of the wake-up instant $i$ such that $\gamma(i) \gteq \varepsilon$, and the second one is for $\gamma(i) \lthan \varepsilon$.

\vspace{-4mm}
\subsection{Message-Delivery Probability for Message Replication}
\label{sec:MR_analysis}
To derive the message delivery probability, consider an arbitrary message $\psi$ from an arbitrary ED $\E$. Suppose $\E$ wakes up in a slot $i$. Then the following events can occur:

% \subsubsection{$\gamma(i) > 0$} 
% Of the $m$ messages, $\hat{\varepsilon}(i)$ are sent twice, and $m \minus \hat{\varepsilon}(i)$ are sent once. Thus, a given message is sent only once with probability $\mathcal{P}_1(i) \equals (m \minus \hat{\varepsilon}(i))/m$ and exactly twice with probability $\mathcal{P}_2(i) \equals \hat{\varepsilon}(i)/m$. If it is sent only once, the message delivery probability is approximately $\hat{\zeta}$; if sent twice, the corresponding approximate value is $1-(1-\hat{\zeta})^2$, where $\hat{\zeta}$ is defined as in~\eqref{eq:avg_zeta}.   
% \begin{align}
%     \hat{\mathcal{S}}_1(i) = \mathcal{P}_1(i) \hat{\zeta} + \mathcal{P}_2(i) [1-(1-\hat{\zeta})^2].
% \end{align}

\subsubsection{$\gamma(i) > 0$} 
Of the $\beta$ messages, $m_r(i)$ are sent $m_q(i) \plus 1$ times, and the remaining $\beta \minus m_r(i)$ are sent $m_q(i)$ times. Thus, $\mathcal{P}_1(i) \equals (\beta \minus m_r(i))/\beta$ and  $\mathcal{P}_2(i) \equals m_r(i)/\beta$ denote the probability that a given message is sent $m_q(i) $ times and $m_q(i) \plus 1$ times, respectively. The corresponding probabilities that at least one copy of the message is delivered are approximately $1-(1-\hat{\zeta})^{m_q(i)}$ and $1-(1-\hat{\zeta})^{m_q(i) \plus 1}$, where $\hat{\zeta}$ is defined as in~\eqref{eq:avg_zeta}.   
\begin{align} \nonumber
    \hat{\mathcal{S}}_1(i) = &\mathcal{P}_1(i) \left( 1-(1-\hat{\zeta})^{m_q(i)} \right) \\
    &+
    \mathcal{P}_2(i) \left( 1-(1-\hat{\zeta})^{m_q(i) + 1} \right).
\end{align}

\subsubsection{$\gamma(i) \leq 0$} 
In this case, 
the probability that a given message is sent over slot $s$, and the subsequent message delivery probability are the same as in the fountain-coding scheme, and are given by~\eqref{eq:tx_prob_2} and \eqref{eq:suc_prob_2}, respectively.

Using a similar argument as used for~\eqref{eq:S_FC}, the overall message delivery probability for message replication is  
\begin{align}
    \hat{\mathcal{S}} = \sum_{i=0}^{N_s-\beta}P_W(i)\hat{\mathcal{S}}_1(i) + \sum_{i=N_s-\beta+1}^{N_s-1}P_W(i)\hat{\mathcal{S}}_2(i),
\end{align}
where $\hat{\mathcal{S}}_2(i)$ is the same as $\mathcal{S}_2(i)$ given by~\eqref{eq:suc_prob_2}.

\section{Numerical Results}
\label{eq:results}

For simulations, we consider 30 EDs distributed over a circular region of 30~m radius, UAV altitude of 10\,m, $\beta \equals 5$, \mbox{$N_f \equals 8$}, $q \equals 256$ for fountain coding,  SF set $\{7,8,9\}$, and Nakagami-$m$ fading with $m \equals 3$ and path-loss exponent 2.5. The simulations are performed in MATLAB. Every data point is obtained by averaging 10,000 simulation runs.

\begin{figure}
    \centering
    \includegraphics[scale=0.42]{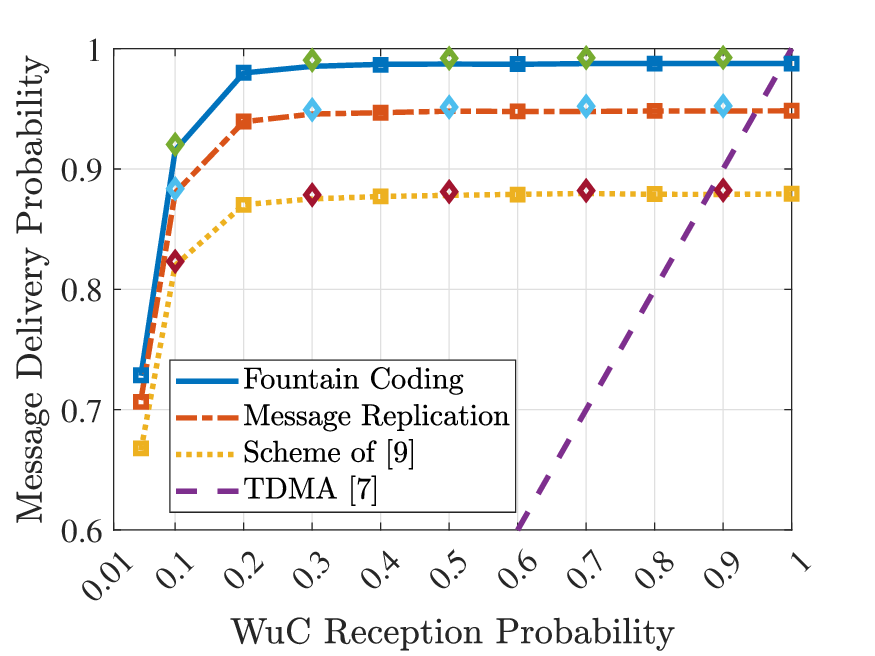}
    \caption{Impact of WuC reception probability \mbox{$(\varepsilon \equals 5$, $N_s \equals 30$).} The lines show simulation results. Squares and diamonds show analytical results that utilize~\eqref{eq:F} and~\eqref{eq:F_AWGN}, respectively, in~\eqref{eq:TSP}.}
    \label{fig:Fig1_vs_Pb_fading}
    \vspace{-5mm}
\end{figure}

Fig.~\ref{fig:Fig1_vs_Pb_fading} shows the message delivery probabilities for \mbox{$\varepsilon \equals 5$} as a function of the WuC reception probability $P_b$. For comparison, we show results for the random-access scheme of~\cite{PaB23}, which does not employ erasure correction. We also compare against a TDMA scheme similar to~\cite{SAA25}. In~\cite{SAA25}, the UAV transmits a WuC when it arrives. EDs that successfully receive the WuC send joining requests through their main radios. The UAV assigns TDMA slots to the sensors from which it receives joining requests. Our evaluation of the TDMA benchmark considers the best-case scenario in which joining-request reception is always successful, that is, any device that receives the WuC is allotted time slots.   We observe that the proposed schemes consistently outperform~\cite{PaB23}, demonstrating the benefits of erasure correction. Also, fountain coding outperforms message replication due to the richer redundancy where every coded message carries information about all source messages as opposed to message replication carrying one source message per redundant frame. The TDMA scheme of~\cite{SAA25} performs poorly when $P_b$ is small (EDs often fail to wake up, and are hence not assigned transmission slots). Only for $P_b \gthan 0.9$, TDMA performs better than the proposed schemes.         

\begin{figure}
    \centering
    \includegraphics[scale=0.42]{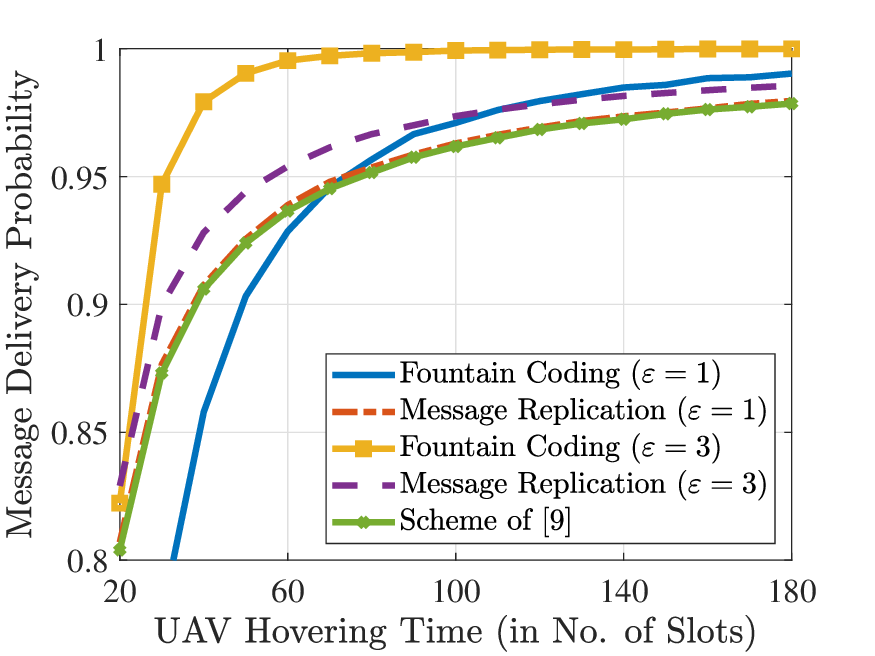}
    \caption{Impact of UAV hovering time ($P_b \equals 0.25$).}
    \label{fig:fig2_twoEPS}
        \vspace{-5mm}
\end{figure}

Fig.~\ref{fig:fig2_twoEPS} shows the impact of UAV hovering time. In all cases, performance improves with hovering time, due to reduced collisions. For $\varepsilon \equals 3$, erasure correction schemes outperform~\cite{PaB23}.  For $\varepsilon \equals 1$, message replication barely provides any improvement over~\cite{PaB23}, since it replicates only one out of the $\beta$ messages, and thus only a loss of this message can be corrected. In addition, fountain coding performs much worse than other schemes when $N_s \lthan 70$; this is because when $\varepsilon \equals 1$, the set of coded messages is not decodable whenever there is more than a single erasure. %Even with a single or no erasure,  decoding may fail due to the rank of the set of received packets being less than $m$; however, with a large $q$ (e.g., as in our case), that probability is quite small. 
For $N_s \gthan 70$, frame losses become rarer, and fountain coding outperforms others. Thus, Fig.~\ref{fig:fig2_twoEPS} illustrates the need for careful protocol selection depending on the redundancy budget and the available contact time.  

\begin{figure}
    \centering
\includegraphics[scale=0.42]{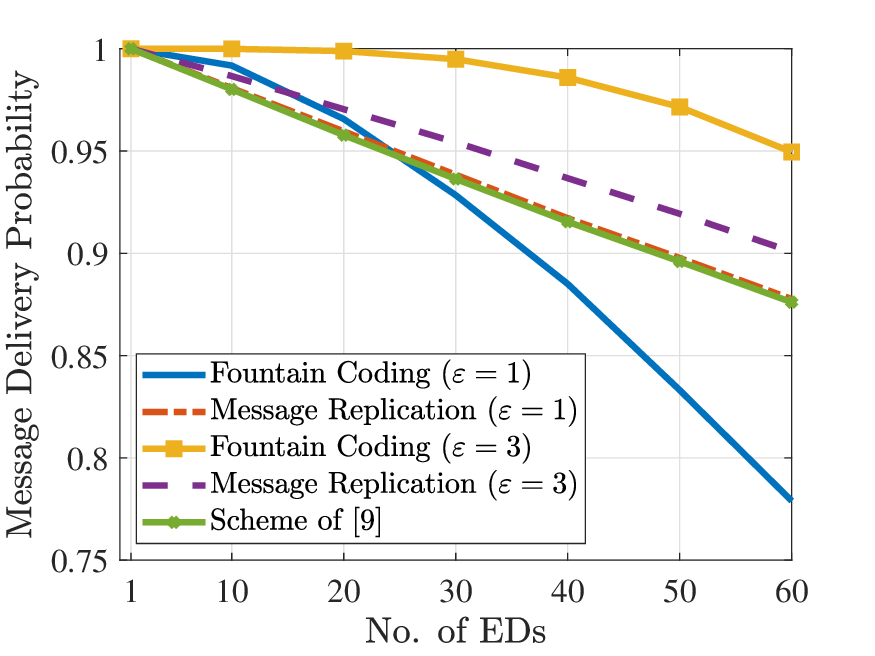}
    \caption{Impact of node density $n$  ($N_s \equals 60, P_b \equals 0.25$).}
    \label{fig:fig3_vsNumEDs}
    \vspace{-6mm}
\end{figure}

Fig.~\ref{fig:fig3_vsNumEDs} shows the impact of the number of EDs.  Since collisions increase with node count, the performance of all scheme degrades. But with $\varepsilon \equals 3$, erasure correction outperforms [9], with fountain coding being significantly better. For $\varepsilon \equals 1$, message replication performs virtually the same as the baseline, whereas fountain coding is worse than the baseline for high node densities (due to more frame losses leading to rank deficiency) and is better when there are fewer EDs.   

%We remark here that although we considered WuR and LoRa for demonstration, the principles can be generalized to any wake up scheme and  

% \vspace{-2mm}
\section{Conclusion}
We described two erasure-correction schemes for data recovery in UAV-IoT data collection. We demonstrated their performance gains over existing approaches and identified how they are impacted by the UAV's hovering time, the maximum amount of redundancy transmission permitted by the sensor's energy budget, and device density. By improving the data-collection performance without a large complexity overhead, these schemes are expected to benefit a wide range of low-altitude UAV-IoT use cases such as precision agriculture and infrastructure monitoring.

\begin{appendices}

\section{Sensor Transmission Limits: Example Calculation}
\label{App:Nmax}

Let $L_f(k)$ be the length of a LoRa frame using SF $k$. Thus, the average duration of a frame is 
$
    \overline{L}_f =  \eta\sum_{k=7}^{K_m}L_f(k)
$, where $\eta \equals 1/(k_m-6)$ is the probability of choosing an SF.  Let $I_t$ be the current consumed by the main radio  while transmitting, $I_c$ be the device's current consumption while performing sensing-cum-computing tasks, and $C_b$ be the battery capacity. Suppose that a device performs sensing and computational tasks for $T_c$ seconds per day and the UAV visits the cluster $V$ times per day. To ensure that the sensor battery lasts approximately $L$ days, the number of frames $N$ the sensor can transmit per UAV visit must satisfy
\begin{align}\label{eq:nmax_ineq}
    (N \overline{L}_f I_t V + T_cI_c)L \leq C_b.   
\end{align} 
The maximum possible value of $N$ is therefore
\begin{align}
    N_{\max}  = \left\lfloor \frac{C_b - LT_cI_c}{LV\overline{L}_f I_t} \right\rfloor.  
\end{align}
For example, suppose $C_b \equals 600$ mAH,  $L \equals 2$ years, $V \equals 12$, and $T_c \equals 20$~s, $k_m \equals 9$, and each message is 50 bytes. Assuming $I_t \equals 83$ mA~\cite{CMV17} and $I_c \equals 50$ mA, we obtain $N_{\max} \equals 10$.

\section{Frame-Loss Probability}
\label{App:FLP}

We will assume that the EDs are uniformly distributed over a circular region with radius $\mathcal{R}$ and the UAV is located $h$ meters above the cluster center. Thus, the pdf for the distance $D$ between an arbitrary ED and the UAV is
\begin{align}
    f_D(u) =
    \begin{cases}
        {2u}/{\mathcal{R}^2}, \quad &h \leq u \leq \sqrt{\mathcal{R}^2+h^2} \\
        0,\quad &\text{otherwise}.
    \end{cases}
\end{align}
and
\begin{align}
    F_D(u) =
    \begin{cases}
        0, \quad &u < h\\
        {(u^2-h^2)}/{\mathcal{R}^2}, \quad &h \leq u \leq \sqrt{\mathcal{R}^2+h^2} \\
        1,\quad &\text{otherwise}.
    \end{cases}
\end{align}
Consider a frame $\F$ from an ED at distance $d_0$. The power received at the UAV is $R = A \gamma_0 p_t d_0^{-\alpha}$, where $p_t$ is the transmit power, $\alpha$ is the path-loss exponent, $\gamma_0$ is a hardware-dependent constant, and $A$ is the power-fading coefficient. For Nakagami-$m$ block fading, $A$ is gamma distributed with its probability and cumulative distribution functions given by 
\begin{align}
    f_A(a) = \frac{1}{\Gamma(m)} \left(\frac{m}{\Omega}\right)^m a^{m-1} \e^{-\frac{ma}{\Omega}}
\end{align}
and
\begin{align}
    F_A(a) = \frac{1}{\Gamma(m)} \gamma\left(m,\frac{ma}{\Omega}\right), \quad a \gthan 0,
\end{align}
respectively, where $\Omega \equals \mathrm{E}[A]$,  $\Gamma(m) \equals \int_0^{\infty} t^{m-1}\e^{-t} dt$, and $\gamma(m,x) \equals \int_0^{x} t^{m-1}\e^{-t} \dx{t}$. Now consider a frame $\F'$ that collides with $\F$ on the same frequency channel. The
received power in $\F'$ is $R' \equals \gamma_0 p_l A' u^{-\alpha}$, where $A'$ is the fading coefficient and $u$ is the distance to the UAV from the sender of $\F'$. Let $k$ and $k'$ be the SFs used by $\F$ and $\F'$, respectively. Assuming the widely used strongest interferer model~\cite{GeR17}, $\F'$ causes a loss of $\F$ if $R/R' \lthan \xi_{k,k'}$, where $\xi_{k,k'}$ is the capture threshold (see~\cite{MSG19} for values). This occurs with probability

\begin{align} \label{eq:power_cond} 
    \mathcal{\tilde{F}}  \nonumber
    &= P(R/R' <  \xi_{k,k'}) \\ \nonumber
    &= P(ad_0^{-\alpha}/A'u^{-\alpha} <  \xi_{k,k'}) \\ \nonumber 
    &= P(A'> \xi_{k,k'}^{-1}a(d_0/u)^{-\alpha}) \\ \nonumber
    &= 1 - F_A( \xi_{k, k'}^{-1}a(d_0/u)^{-\alpha}) \\ 
    &= 1 - \frac{1}{\Gamma(m)} \gamma\left(m,m \xi_{k, k'}^{-1}a(d_0/u)^{-\alpha}\right).
\end{align}

Removing the conditioning on $k$, $k'$, $a$, $d_0$, and $u$, we obtain
\begin{align} \label{eq:F}
    \mathcal{F} 
    &\equals \frac{1}{\eta^2}\sum_{k=7}^{k_m} \sum_{k'=7}^{k_m}\displaystyle\int_{h}^{w}\int_{h}^{w} \int_{0}^{\infty} \tilde{\mathcal{F}} f_A(a) f_D(d_0) f_D(u) \dx{a} \dx{d_0} \dx{u},
\end{align}
where $w \equals \sqrt{\mathcal{R}^2 \plus h^2}$ and $1/\eta^2 \equals 1/(k_m \minus 6)^2$ is the probability of choosing SFs $k$ and $k'$, respectively, for the desired and colliding frames. 

\paragraph{A Closed-Form Approximation for $\mathcal{F}$}  A simple expression for $\mathcal{F}$ can be derived by ignoring fading on the links, that is, letting $R \equals \gamma_0 p_t d_0^{-\alpha}$ and $R' \equals \gamma_0 p_t u^{-\alpha}$ and evaluating
\begin{align} \label{eq:power_cond} 
    \mathcal{\tilde{F}}(u)  \nonumber
    &\approx P(\gamma_0p_td_0^{-\alpha}/\gamma_0p_tu^{-\alpha} <  \xi_{k,k'}) \\ \nonumber 
    &= P(d_0> \xi_{k,k'}^{-1/\alpha}u) \\ 
    &= \begin{cases}
    1, \hspace{37mm}u < \xi_{k,k'}^{1/\alpha}h\\ 
    1-(\xi_{k,k'}^{-2/\alpha}u^2 - h^2)/\mathcal{R}^2, \quad \xi_{k,k'}^{1/\alpha}h \leq u \leq \xi_{k,k'}^{1/\alpha}w\\
    0, \hspace{37mm}u > \xi_{k,k'}^{1/\alpha}w,
    \end{cases}
    \end{align}
where $w \equals \sqrt{\mathcal{R}^2+h^2}$. Removing the conditioning on $k'$, $k$, and $u$, we obtain the following approximation to~\eqref{eq:F}
\begin{equation} \label{eq:F_AWGN} 
    \mathcal{F} 
    \approx \frac{1}{\eta^2}\sum_{k=7}^{k_m} \sum_{k'=7}^{k_m}\int_{h}^{w}\tilde{\mathcal{F}}(u)f_D(u)\dx{u} = \frac{1}{\eta^2}\sum_{k=7}^{k_m} \sum_{k'=7}^{k_m}I,
\end{equation}
where the integral $I$ evaluates to
\begin{align}
    I \equals \begin{cases}
    0, \quad &0 < \xi_{k,k'}^{1/\alpha} \lthan \frac{h}{w}\\
    (b^2-h^2)c - \xi_{k,k'}^{-2/\alpha}(b^4-h^4), \quad &\frac{h}{w} \lthan \xi_{k,k'}^{1/\alpha} \lthan 1\\
    \frac{a^2-h^2}{\mathcal{R}^2} + (w^2 \minus h^2)c - \frac{\xi_{k,k'}^{-2/\alpha}}{2\mathcal{R}^2}(w^4 \minus a^4), \quad &1 \lthan \xi_{k,k'}^{1/\alpha} \lthan \frac{w}{h}\\
    1, \quad &  \frac{w}{h} \lthan \xi_{k,k'}^{1/\alpha} \lthan w,
    \end{cases}
\end{align}

where $a \equals \xi_{k,k'}^{1/\alpha}h$, $b \equals \xi_{k,k'}^{1/\alpha}w$, and $c \equals 1/\mathcal{R}^2 \plus h^2/\mathcal{R}^4$.

\end{appendices}

\vspace{-2mm}
\bibliographystyle{ieeetr}
% \bibliography{IEEEabrv,refs.bib}
\bibliography{refs.bib}

\end{document}